\documentclass[prl,twocolumn,superscriptaddress]{revtex4}
\usepackage{graphicx}

\input{tcilatex}

\begin{document}

\title{Resonant Spin Hall Conductance in Two-Dimensional Electron Systems
with Rashba Interaction in a Perpendicular Magnetic Field}
\author{Shun-Qing Shen$^{1}$, Michael Ma$^{2}$, X.C Xie$^{3,4}$, and Fu Chun
Zhang$^{1,2,5}$}
\affiliation{$^{1}$Department of Physics, The University of Hong Kong, Pukfulam Road,
Hong Kong, China\\
$^{2}$Department of Physics, University of Cincinnati, Ohio 45221\\
$^{3}$Department of Physics, Oklahoma State University, Stillwater, Oklahoma
74078\\
$^{4}$ICQS, Institute of Physics, Chinese Academy of Sciences, Beijing,
China \\
$^{5}$Department of Physics, Zhejiang University, Hangzhou, Zhejiang, China }
\date{February 23, 2004}

\begin{abstract}
We study transport properties of a two-dimensional electron system with
Rashba spin-orbit coupling in a perpendicular magnetic field. The spin orbit
coupling competes with Zeeman splitting to introduce additional degeneracies
between different Landau levels at certain magnetic fields. This degeneracy,
if occuring at the Fermi level, gives rise to a resonant spin Hall
conductance, whose height is divergent as $1/T$ and whose weight is
divergent as $-\ln T$ at low temperatures. The Hall conductance is
unaffected by the Rashba coupling..
\end{abstract}

\maketitle

Remarkable phenomena have been observed in the two-dimensional electron gas
(2DEG) over last two decades, including most notably, the discoveries of the
integer and fractional quantum Hall effect \cite{Klitzing80,Tsui82}. From
the point of view of applications, many semiconductor devices have been
designed to take advantage of the properties of quantum physics.
Nevertheless, a principal quantum aspect of an electron, its spin, has been
largely ignored. In recent years, however, a new class of devices based on
the spin degrees of freedom of electrons has emerged, giving rise to the
field of spintronics.\cite{Prinz98Science,Wolf01Science,Awschalom02}
Spintronics is believed to be a promising candidate for future information
technology.\cite{Loss98} However, in order to be successful in device
applications, effective spin injection into conventional semiconductors is
essential. One proposal is to make use of the Rashba spin-orbit coupled
2DEGs to achieve this goal.\cite{Datta90} In particular, the spin-Hall
effect predicted by Murakami \textit{et al} \cite{Murakami03Science} and
Sinova \textit{et al} \cite{Sinova03xxx} has generated intensive theoretical
studies. Thus far, all the studies have been limited to zero magnetic field.%
\cite{Shen03xxx}

In this Letter, we study theoretically the spin transport properties of
2DEGs with a Rashba spin-orbit coupling in a perpendicular magnetic field.
We find that the quantized charge Hall conductance remains intact in the
presence of the Rashba spin-orbit coupling. However, a distinct spin Hall
current can be generated. The spin Hall conductance can be made divergent or
resonant by tuning the sample parameters and/or magnetic field $B$. The
resonance effect stems from energy crossing of different Landau levels near
the Fermi level due to the competition of Zeeman energy splitting and
spin-orbit coupling. The height of the resonant peak in spin Hall
conductance is proportional to $1/T$, and its weight is proportional to $%
-\ln T$ at low temperatures. 
\begin{figure}[tbp]
\includegraphics*[width=8cm, angle=0]{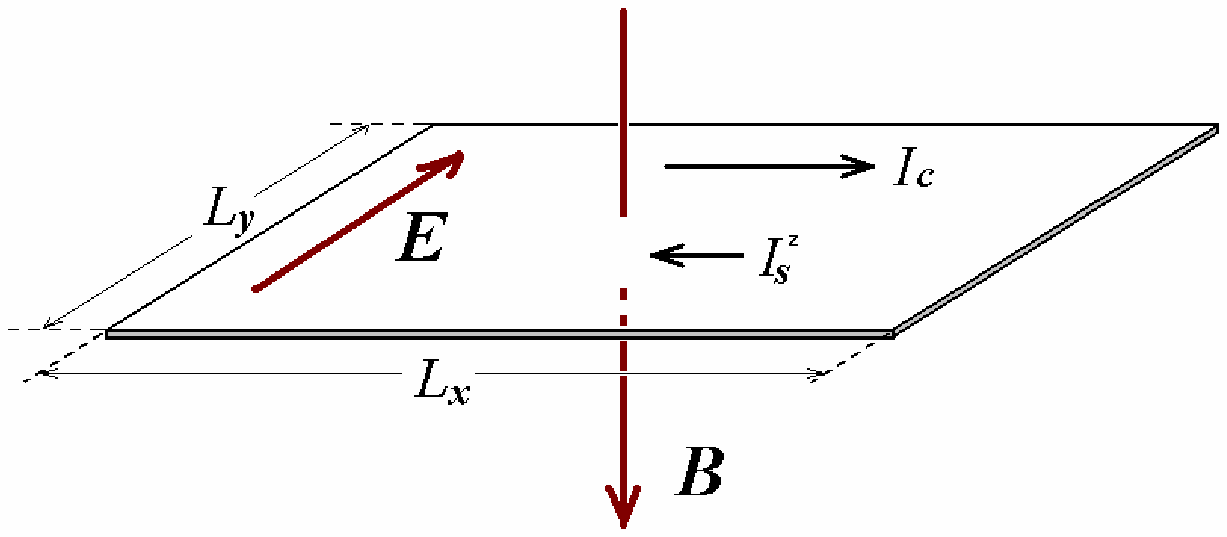}
\caption{Illustration of the two dimensional electron system studied in the
text. }
\end{figure}

We consider a two-dimensional electron system confined in the $x-y$ plane of
an area $L_{x}\times L_{y}$ provided by a semiconductor quantum well as
shown in Fig.1. The electron is subject to a spin-orbit interaction and to a
perpendicular magnetic field $\vec{B}=-B\hat{z}$. An electric field is
applied along the $y-$ axis. We are interested in the spin Hall conductance
along the $x-$ direction. In our study, electron-electron interactions are
neglected. The Hamiltonian for a single electron of spin-1/2 is given by

\begin{eqnarray}
H &=&\frac{1}{2m}(\vec{p}+\frac{e\vec{A}}{c})^{2}+\frac{\lambda }{\hbar }%
\hat{z}\cdot (\vec{p}+\frac{e\vec{A}}{c})\times \vec{\sigma}  \nonumber \\
&&-\frac{1}{2}g_{s}\mu _{b}B\sigma _{z}+eEy,
\end{eqnarray}%
where $m$, $-e$, $g_{s}$ are the electron's effective mass, charge and
Lande- $g$ factor, respectively. $\mu _{b}$ is the Bohr magneton, $\lambda $
is the Rashba coupling, and $\sigma _{\gamma }$ are the Pauli matrices. We
choose the Landau gauge $\vec{A}=yB\hat{x}$, and consider periodic boundary
condition in the $x$-direction, hence $p_{x}=k$ is a good quantum number.

\begin{figure}[tbp]
\includegraphics*[width=8.7cm,height=6.2cm,angle=0]{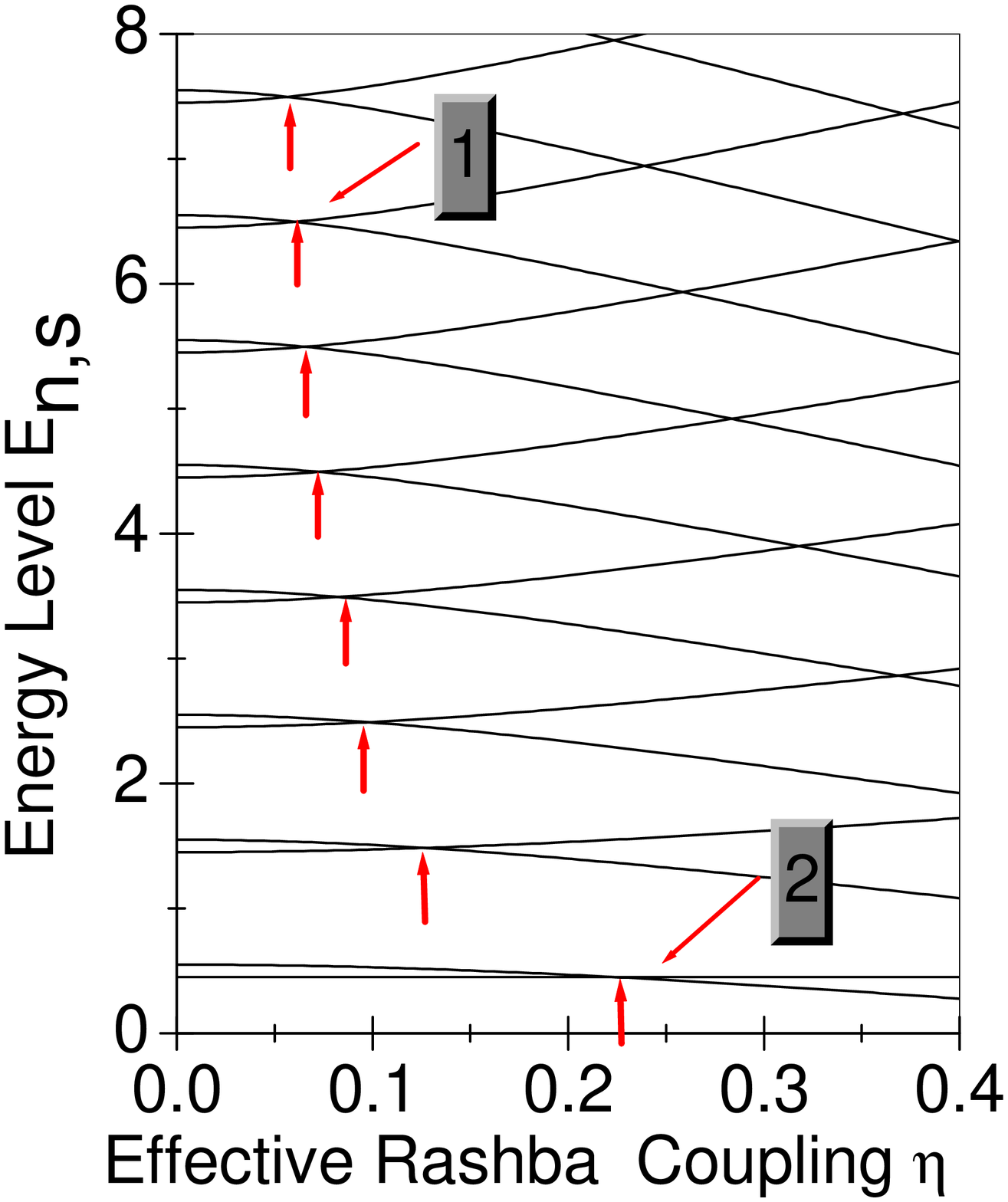}
\caption{Landau levels of an electron as functions of $\protect\eta =\protect%
\lambda ml_{b}/\hbar ^{2}$ for $g=g_{s}m/2m_{e}=0.1$. Arrows indicate those
level crossings giving rise to resonant spin Hall conductance. }
\end{figure}

Let us start with a discussion of the single particle solution at $E=0$. The
problem can be solved exactly. \cite{Rashba60,Luo88,Schliemann03} For a
given k, the Hamiltonian can be written as, 
\begin{equation}
H_{0}=\hbar \omega \{a_{k}^{\dag }a_{k}+\frac{1-g\sigma _{z}}{2}+\sqrt{2}%
\eta (ia_{k}\sigma _{-}-ia_{k}^{\dag }\sigma _{+})\},  \label{h0}
\end{equation}%
\newline
where $\omega =eB/mc$, $\eta =\lambda ml_{b}/\hbar ^{2}$, and $%
g=g_{s}m/2m_{e}$, with $m_{e}$ the mass of a free electron and $l_{b}=\sqrt{%
\hbar c/eB}$ the magnetic length. $a_{k}=[y+(k+ip_{y})c/eB]/\sqrt{2}l_{b}$,
so that $[a_{k},a_{k}^{\dag }]=1$. $\sigma _{\pm }=(\sigma _{x}\pm \sigma
_{y})/2.$ The eigen energy of $H_{0}$ is given by 
\begin{equation}
\epsilon _{ns}=\hbar \omega \left( n+\frac{s}{2}\sqrt{(1-g)^{2}+8n\eta ^{2}}%
\right) \,,
\end{equation}%
with $s=\pm 1$, for $n\geq 1$; and $s=1$ for $n=0$. The eigenstate $%
\left\vert n,k,s\right\rangle $ has a degeneracy $N_{\phi }=L_{x}L_{y}eB/hc$%
, corresponding to $N_{\phi }$ quantum values of $k$. The two-component
wavefunction is given by

\begin{equation}
\left\vert n,k,s\right\rangle =\left( 
\begin{array}{c}
\cos {\theta _{ns}}\phi _{nk} \\ 
i\sin {\theta _{ns}}\phi _{n-1k}%
\end{array}%
\right)
\end{equation}%
where $\phi _{nk}$ is the eigenstate of the $n^{th}$ Landau level in the
absence of the Rashba interaction. For $n=0,$ $\theta =0$, otherwise for $%
n\geq 1$, $\tan {\ \theta _{ns}}=u_{n}-s\sqrt{1+u_{n}^{2}}$, with $%
u_{n}=(1-g)/\sqrt{8n}\eta $. The energy levels as functions of dimensionless
parameter $\eta $ are plotted in Fig. 2. An interesting feature of this
system is the energy level crossing as $\eta $ changes by varying $B$ or $%
\lambda $. As we shall see below, this energy crossing, if it occurs at the
Fermi level, gives rise to a resonance in the spin Hall conductance.

We now study the system in the presence of the $E$-field. The Hamiltonian $H$
can be rewritten as 
\begin{eqnarray}
H &=&H_{0}(E)+H^{\prime },  \nonumber \\
H^{\prime } &=&-(\eta el_{b}\sigma _{y}+kc/B)E,
\end{eqnarray}%
where we have dropped an overall constant $-e^{2}E^{2}/2m\omega ^{2}$. $%
H_{0}(E)$ is given by Eq. (\ref{h0}) of $H_{0}$ with the replacement of $y$
by $y+eE/m\omega ^{2}$ in $a_{k}$. In the absence of Rashba coupling, $H$ is
exactly solvable. For $\lambda \neq 0$ and $E\neq 0$, an exact solution of $%
H $ is not available. While $p_{x}$ remains to be a good quantum number, $%
H^{\prime }$ couples the state $\left\vert n,k,s\right\rangle $ with $%
\left\vert n\pm 1,k,s^{\prime }\right\rangle $. Below we shall calculate the
charge and spin Hall current to the order of $O(E)$ by treating $H^{\prime }$
as a perturbation up to the first order. Our theory is accurate for the
linear response. The charge current operator of a single electron is given
by \cite{currentdef} 
\begin{eqnarray}
j_{c} &=&-ev_{x},  \nonumber \\
v_{x} &=&[x,H]/\left( i\hbar \right) =p_{x}/m+y\omega +(\lambda /\hbar
)\sigma _{y},
\end{eqnarray}%
and the spin-$\gamma $ component current operator is 
\begin{equation}
j_{s}^{\gamma }=\frac{\hbar }{2}(S^{\gamma }v_{x}+v_{x}S^{\gamma }).
\end{equation}%
Let $(j_{c,s})_{nks}$ be the current carried by an electron in the state $%
\left\vert n,k,s\right\rangle $ of $H$, including also the perturbative
correction. We have, up to the first order in $E$, 
\begin{equation}
(j_{c,s})_{nks}=(j_{c,s}^{(0)})_{nks}+(j_{c,s}^{(1)})_{nks},
\end{equation}%
where the superscript refers to the $0^{th}$ or $1^{st}$ order in the
perturbation in $H^{\prime }$, and 
\begin{eqnarray}
(j_{c,s}^{(0)})_{nks} &=&\left\langle n,k,s\right\vert j_{c,s}\left\vert
n,k,s\right\rangle ,  \nonumber \\
(j_{c,s}^{(1)})_{nks} &=&\sum_{n^{\prime }s^{\prime }}\frac{\left\langle
n^{\prime },k,s^{\prime }\right\vert H^{\prime }\left\vert
n,k,s\right\rangle \left\langle n,k,s\right\vert j_{c,s}\left\vert n^{\prime
},k,s^{\prime }\right\rangle }{(\epsilon _{ns}-\epsilon _{n^{\prime
}s^{\prime }})}  \nonumber \\
&&+h.c..
\end{eqnarray}%
In the above equation, $n^{\prime }=n\pm 1$ since the matrix element
vanishes for other values of $n^{\prime }$. Note that $H_{0}(E)$ depends on $%
E$ so that the $0^{th}$ order in $H$ also contributes to the current. The
average current density of the $N_{e}$-electron system is given by 
\begin{equation}
I_{c,s}=\frac{1}{L_{y}}\sum_{nks}(j_{c,s})_{nks}f(\epsilon _{nks}),
\end{equation}%
where $f$ is the Fermi distribution function, and $N_{e}=\sum_{nks}f(%
\epsilon _{nks})$. The charge or spin Hall conductance is then given by $%
G_{c,s}=I_{c,s}/E$.

\begin{figure}[tbp]
\includegraphics*[width=8.7cm,height=6.2cm,angle=0]{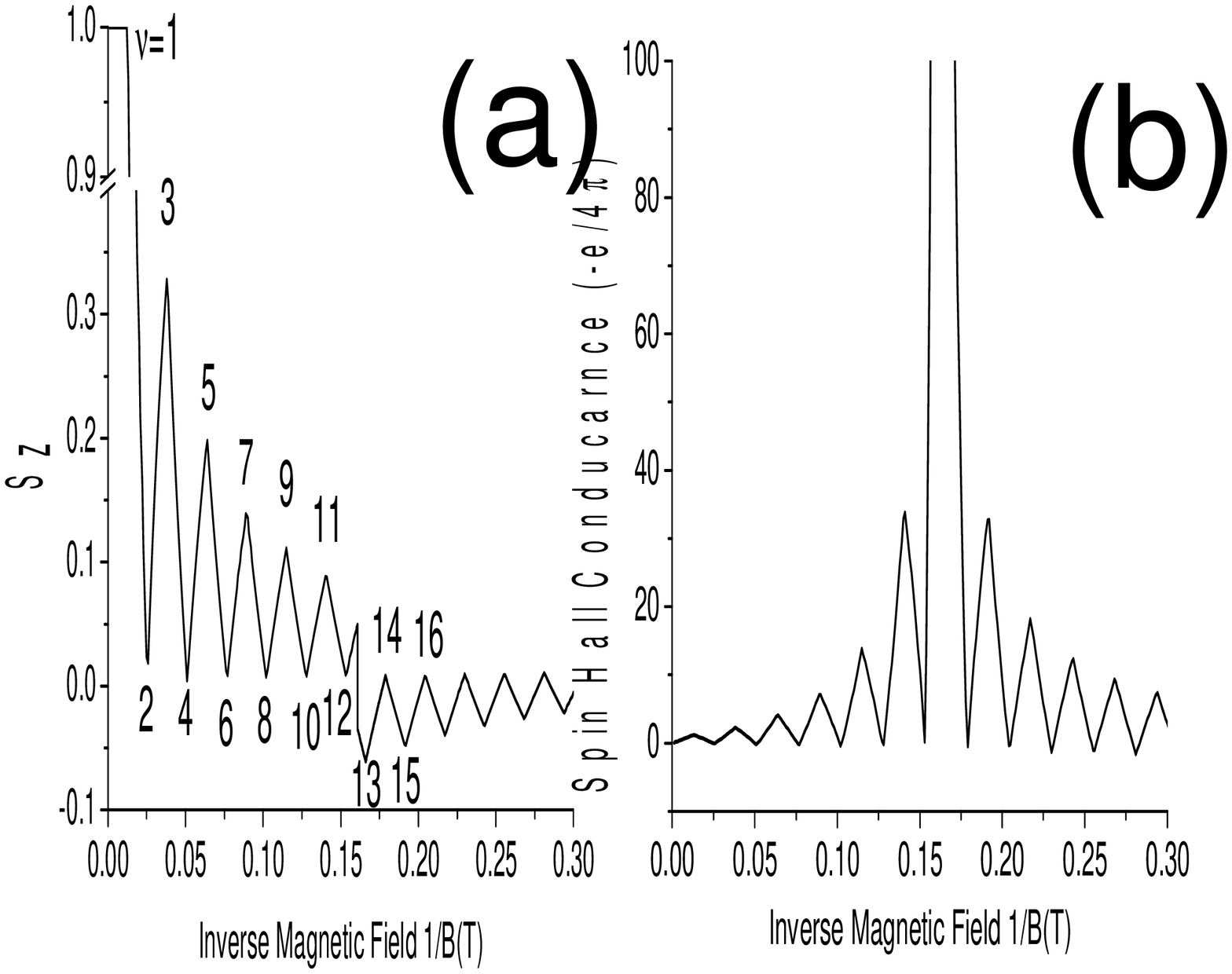}
\caption{Left: Average spin $\left\langle \protect\sigma _{z}\right\rangle $
(unit: $\hbar /2$) per electron as a function of $1/B$. Right: spin Hall
conductance versus $B$ at $T=0$. The parameters used are $\protect\lambda %
=0.9\times 10^{-11}eVm,$ $n_{e}=1.9\times 10^{16}/m^{2}$, $g_{s}=4$, and $%
m=0.05m_{e}$, taken for the inversion heterostructure In$_{0.53}$Ga$_{0.47}$%
As/In$_{0.52}$Al$_{0.48}$As. (See Ref.\protect\cite{Nitta97prl}).}
\end{figure}

We first present the results for the charge current and the spin currents in
the spin-$x$ and -$y$ components. They are found to be 
\begin{eqnarray}
(j_{c})_{nks} &=&(EL_{y})e^{2}/hN_{\phi },  \nonumber \\
(j_{s}^{x})_{nks} &=&0,  \nonumber \\
(j_{s}^{y})_{nks} &=&\frac{\lambda }{2\hbar }\left( 1+\frac{s}{\eta }\sqrt{%
\frac{n}{2(1+u_{n}^{2})}}\right) \,.
\end{eqnarray}%
From the above expression, we obtain that the Hall conductance $G_{c}=\nu
e^{2}/h$, with the filling factor $\nu =N_{e}/N_{\phi }$. In fact this
result holds to all order in $E$. This is because the only k-dependence of
the energy comes from the $-kcE/B$ term in $H$, thus the group velocity $%
v_{x}=cE/B$. This result is consistent with the quantization of the Hall
conductance:\cite{Prange87} \textit{the spin-orbit coupling does not change
the charge current carried by each state}. The spin-$y$ component current is
found to be finite even in the absence of $E$. Similar result was reported
previously in the systems at $B=0$.\cite{Rashba03prb} In the limit $%
B\rightarrow 0$, our result gives $I_{s}^{y}\rightarrow \lambda
n_{e}/(2\hbar )$, which is the same as the result found at $B=0$. Since this
is not a response to any external field, we will not give further discussion
here.

The spin-z component current is the most interesting. Within the
perturbation theory, $I_{s}^{z}$, hence $G_{s}^{z},$ can be divided into two
parts. The part arising from the $0^{th}$ order in $H^{\prime }$ is found to
be the product of the spin polarization $S^{z}$ and the Hall conductance $%
G_{c}$ divided by the electron charge ($-e$), 
\begin{eqnarray}
G_{s}^{z}{}^{(0)} &=&-\langle S^{z}\rangle (G_{c}/e),  \nonumber \\
\langle S^{z}\rangle &=&(\hbar /2\nu )\sum_{n,s}\cos {(2\theta _{ns})}%
f(\epsilon _{ns})
\end{eqnarray}%
Since the charge current is a constant, $G_{s}^{z}{}^{(0)}\propto \langle
S^{z}\rangle $. The spin polarization per electron at $T=0$ as a function of
the Landau level filling is plotted in Fig. 3a, for a set of parameters
appropriate for In$_{0.53}$Ga$_{0.47}$As/In $_{0.52}$Al$_{0.48}$As.\cite%
{Nitta97prl}. $\langle S^{z}\rangle $ oscillates as a result of the
alternative occupation of mostly spin-up and mostly spin-down electrons. It
reaches maxima at filling $\nu =$ odd integers, and minima at $\nu =$ even
integers at a strong field $B$. There is a jump at $B=B_{0}\approx 6.1T$ or $%
\nu =12.6$. Below the field, $\langle S^{z}\rangle $ reaches minima at
filling $\nu =$even integers and minima at $\nu =$odd integer. The jump is
caused by the energy crossing of two Landau levels with almost opposite
spins. This value of the filling factor corresponds to the parameter $\eta $
at point $1$ in Fig.2. In the weak field limit, $\langle S^{z}\rangle
\rightarrow -geB/4\pi c,$ and the spin susceptibility approaches to a
constant, $d\langle S^{z}\rangle /dB=-ge/4\pi c.$

The second part in $G_{s}^{z}$ arises from $H^{\prime }$, and shows a
resonance.%
\begin{eqnarray}
&&G_{s}^{z}{}^{(1)}=\frac{e\eta }{8\pi \sqrt{2}}\sum_{n,s,n^{\prime
}=n+1,s^{\prime }}\frac{f(\epsilon _{ns})-f(\epsilon _{n^{\prime }s^{\prime
}})}{(\epsilon _{ns}-\epsilon _{n^{\prime }s^{\prime }})}  \nonumber \\
&&\times \left( \sqrt{n}\sin 2\theta _{ns}\sin ^{2}\theta _{n^{\prime
}s^{\prime }}-\sqrt{n^{\prime }}\cos ^{2}\theta _{ns}\sin 2\theta
_{n^{\prime }s^{\prime }}\right)  \label{gz}
\end{eqnarray}%
Resonance occurs when two states are close to degeneracy. For reasonable
values of the Rashba coupling, this will happen only for the pair of states $%
\left\vert n,s=1\right\rangle $ and $\left\vert n+1,s^{\prime
}=-1\right\rangle .$ However, at $T=0,$ if $\left\vert n,s\right\rangle $
and $\left\vert n+1,s^{\prime }\right\rangle $ are both occupied or both
unoccupied, the contributions to $G_{s}$ from this pair vanish. Therefore,
only the states near the Fermi level are important in the sum in Eq. (\ref%
{gz}). If the two states at the Fermi energy become degenerate, $G_{s}^{z}$
becomes divergent. Therefore, \textit{there is a resonance in the spin Hall
conductance}. The resonant condition (in the clean limit) is given by 
\begin{equation}
\sqrt{(1-g)^{2}+8n\eta ^{2}}+\sqrt{(1-g)^{2}+8\left( n+1\right) \eta ^{2}}=2,
\label{resonance}
\end{equation}%
where $n\leq \nu /2\leq n+1$. In a sample of given $n_{e}$, and $\lambda
\neq 0$ and $1>g>0$, there is a resonant magnetic field $B_{0}$ for the
resonance as the solution of Eq. (\ref{resonance}). In Fig. 3b, we show the
result of $G_{s}^{z}$ at $T=0$ as a function of $\nu $, or $1/B$. In
addition to the oscillations similar to $\langle S^{z}\rangle $, there is a
pronounced resonance at $B_{0}$ or at filling $\nu =12.6$. At this filling
the $13^{th}$ Landau level is partially filled. From Fig. 3b, we also see
that there are satellite peaks around the resonant field $B_{0}$. The
resonance point coincides with the jump point for $\left\langle
S^{z}\right\rangle .$ The spin Hall conductance becomes divergent while $%
\left\langle S^{z}\right\rangle $ has only a finite jump at the energy
crossing point near the Fermi level.

In order to analyze this resonance further, we focus on the two relevant
states and neglect all other states in the problem. For simplicity we
consider the two states $\left\vert 0,+1\right\rangle $ and $\left\vert
1,-1\right\rangle $. The linear response of the two level problem to the
electric field can be studied analytically. The singular part of the spin
Hall conductance near the resonant point (point 2 in Fig. 2) is caused by
the mixing of the two states, and is given by (for filling $\nu <1$), 
\begin{equation}
G_{s}^{z}=-\frac{D\,f_{-}(1-f_{+})}{\left\vert b\right\vert }\left( 1-\exp %
\left[ -\frac{g\hbar \omega _{0}\left\vert b\right\vert }{(1+g)k_{B}T}\right]
\right) ,
\end{equation}%
where $D=e\sqrt{2}\nu g/4\pi (1+g)$, $\omega _{0}$ is the value of $\omega $
at the resonant field $B_{0}$. The Fermi distribution $f_{\pm }=(\exp
[\left( \pm g\hbar \omega _{0}\left\vert b\right\vert /2(1+g)-\mu \right)
/k_{B}T]+1)^{-1}$, where $\mu $ is the chemical potential measured relative
to the mid of the two levels, and $b=(B-B_{0})/B_{0}$. At low $T$, as $%
b\rightarrow 0$, $G_{s}^{z}\rightarrow -Dg(1+g)^{-1}\hbar \omega _{0}/kT$,
and $\int G_{s}^{z}db\rightarrow -D\ln \left[ \hbar \omega _{0}/k_{B}T\right]
$. In Fig. 4, we show $G_{s}^{z}$ (including both singular and non-singular
parts) as a function of $B$ at several temperatures. As we can see, both the
height and the weight of the resonant peak increase as the temperature
decreases.

The calculations reported in this paper have been performed on a 2DEG
without potential disorder. Since the effects of disorder in systems with
Rashba coupling and strong magnetic field is not well understood at this
point, we will make only a few general comments here. We assume that, just
as in the case without Rashba coupling, the presence of disorder gives rise
to broadening of the Landau level and localization so that the extended
states in a Landau levels are separated in energy from those in the next one
by localized states. Inspection of the Rahsba Hamiltonian shows that
Lauhglin's gauge invariant argument still holds,\cite{Laughlin81} and each
Landau level with its extended states completely filled contribute $e^{2}/h$
to the Hall conductance. Thus we conclude that identical quantum Hall effect
is observed whether the Rashba coupling is present or not. For the spin Hall
conductance, we further assume that there is only one extended state per
Landau level as in the case of no Rashba coupling, and that the spin current
is carried only by extended states. The resonance discussed above will then
occur if the extended state of the band $\left\vert n,s\right\rangle $ and
the $\left\vert n+1,s\prime \right\rangle $ band can become degenerate. In
principle, such a degeneracy is disallowed due to level crossing avoidance.
However, since potential disorder does not couple states of different spins,
any coupling between these two states will have to arise from Landau level
mixing effect of the disorder in the absence of Rashba coupling. Provided
this is negligible, the crossing avoidance gap will also be negligible.

In summary, we have studied the transport properties of two dimensional
electron gas with a Rashba spin-orbit coupling in a perpendicular magnetic
field. The Rashba spin-orbit coupling competes with the Zeeman energy
splitting to cause the energy level crossing. When the level crossing occurs
near the Fermi level, the spin Hall conductance becomes divergent or
resonant, while the charge Hall conductance remain intact.

\begin{figure}[tbp]
\includegraphics*[width=6.7cm,height=6.2cm,angle=0]{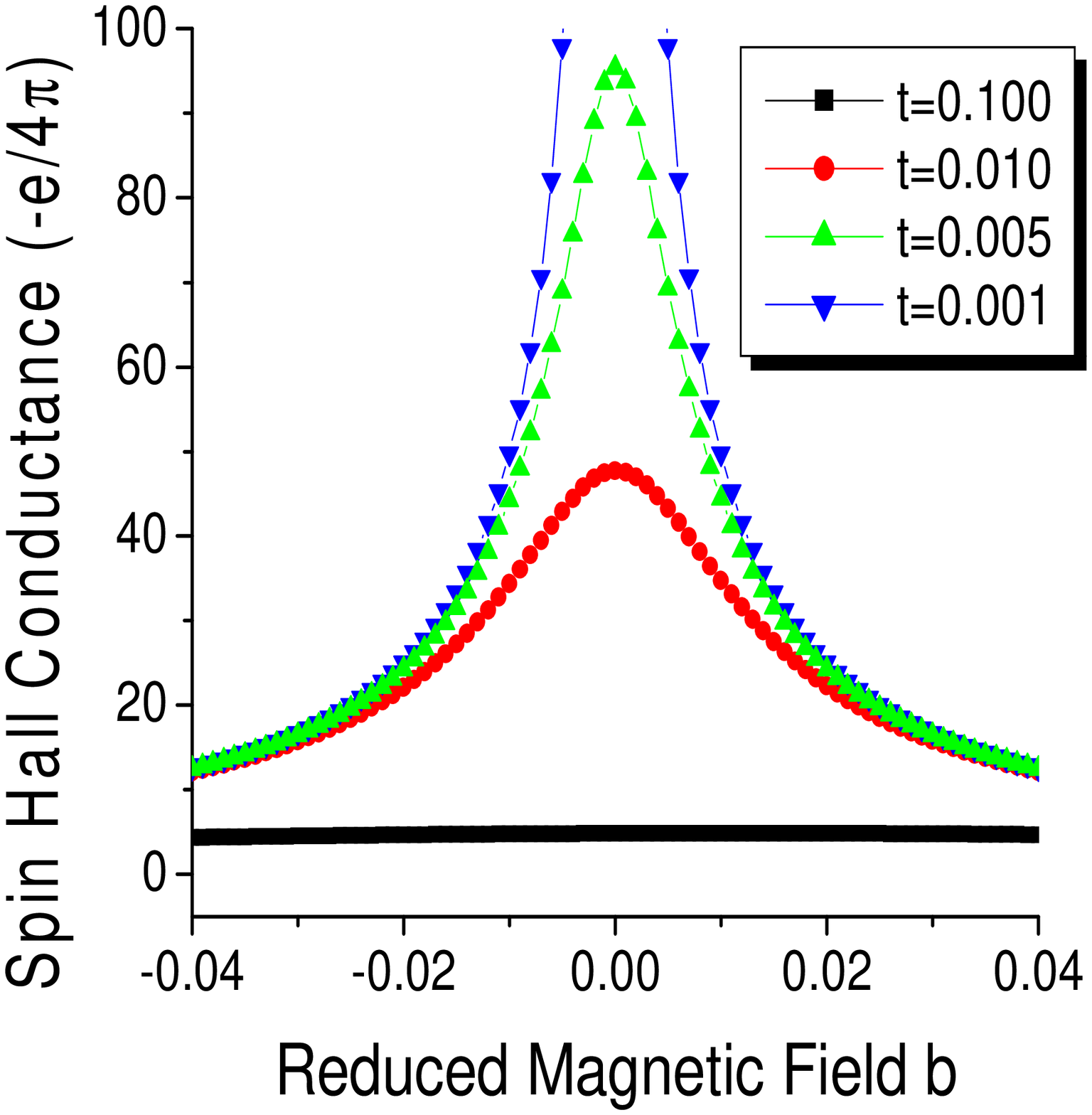}
\caption{Spin Hall conductance as a function of $B$ at various temperatures
for a two level system with $n_{e}$ fixed, $g=0.1,$ and $\protect\nu =0.5$
at the resonant field $B_{0}$. $b=(B-B_{0})/B_{0},$ and $t=kT/\hbar \protect%
\omega (B_{0})$. }
\end{figure}

This work was in part supported by RGC in Hong Kong (SQS), NFC (FCZ), and
DOE/DE-FG02-04ER46124 (XCX).


\begin{thebibliography}{99}
\bibitem{currentdef} For brevity we write the current operators here in the
Hilbert subspace $p_{x}=k$.
\end{thebibliography}
\end{document}